\documentclass[conference]{IEEEtran}
\IEEEoverridecommandlockouts
\usepackage{cite}
\usepackage{amsmath,amssymb,amsfonts}
\usepackage{algorithmic}
\usepackage{graphicx}
\usepackage{textcomp}
\usepackage{xcolor}
\usepackage{hyperref}
\usepackage{colortbl}

\usepackage{booktabs}
\usepackage{textcomp}
\usepackage{stfloats}
\usepackage{url}
\usepackage{verbatim}
\usepackage{multirow}
\usepackage{makecell}
\usepackage{pifont}
\usepackage{threeparttable}
\usepackage{circledsteps}

\definecolor{dark-green}{HTML}{006400}
\definecolor{dark-blue}{HTML}{1976D2}
\definecolor{dark-purple}{HTML}{8d4bbb}
\definecolor{dark-red}{HTML}{D63C3C}
\definecolor{n1}{HTML}{ff9999}
\definecolor{n2}{HTML}{FFCC99}
\definecolor{n3}{HTML}{FFFF99}

\hypersetup{
    colorlinks=true,
    linkcolor=black,
    filecolor=magenta,      
    urlcolor=black,
    citecolor=dark-blue,
}
\newcommand{\shh}[1]{\textcolor{black}{#1}}
\newcommand{\wdm}[1]{\textcolor{black}{#1}}

\usepackage{url}
\def\BibTeX{{\rm B\kern-.05em{\sc i\kern-.025em b}\kern-.08em
    T\kern-.1667em\lower.7ex\hbox{E}\kern-.125emX}}
\begin{document}

\setlength{\abovecaptionskip}{0.5em}

\title{An FPGA-Based Reconfigurable Accelerator for\\ Convolution-Transformer Hybrid EfficientViT
\thanks{This work was supported in part by the National Key R\&D Program of China under Grant 2022YFB4400604. (\textit{Corresponding Author: Wendong Mao and Zhongfeng Wang)}}
}

\author{\IEEEauthorblockN{Haikuo Shao$^{1}$, Huihong Shi$^{1}$, Wendong Mao$^{2}$, and Zhongfeng Wang$^{1,2}$}
\IEEEauthorblockA{$^{1}$School of Electronic Science and Engineering, Nanjing University, Nanjing, China \\
$^{2}$School of Integrated Circuits, Sun Yat-sen University, Shenzhen, China \\
Email: \{hkshao, shihh\}@smail.nju.edu.cn, maowd@mail.sysu.edu.cn, zfwang@nju.edu.cn}
}

\maketitle

\begin{abstract}
Vision Transformers (ViTs) have achieved significant success in computer vision. However, their intensive computations and massive memory footprint challenge ViTs' deployment on embedded devices, calling for efficient ViTs. Among them, EfficientViT, the state-of-the-art one, features a Convolution-Transformer hybrid architecture, enhancing both accuracy and hardware efficiency. Unfortunately, existing accelerators cannot fully exploit the hardware benefits of EfficientViT due to its unique architecture. 
In this paper, we propose an FPGA-based accelerator for EfficientViT to advance the hardware efficiency frontier of ViTs. Specifically, we design a reconfigurable architecture to efficiently support various operation types, including lightweight convolutions and attention, boosting hardware utilization. Additionally, we present a time-multiplexed and pipelined dataflow to facilitate both intra- and inter-layer fusions, reducing off-chip data access costs. Experimental results show that our accelerator achieves up to 780.2 GOPS in throughput and 105.1 GOPS/W in energy efficiency at 200MHz on the Xilinx ZCU102 FPGA, which significantly outperforms prior works.
\end{abstract}


\section{Introduction}
Recently, Vision Transformers (ViTs) have been proposed and attracted increasing attention in the computer vision field \cite{vit, deit}. 
Despite ViTs' remarkable performance against their convolution-based counterparts, the intensive computations and huge memory footprint during inference pose challenges to ViTs' deployment on resource-constrained devices \cite{You2022ViTCoDVT, Dass2022ViTALiTyUL}.
Particularly, the computational complexity of the self-attention mechanism in standard ViTs is quadratic w.r.t. the number of input tokens, limiting ViTs' real-world application on high-resolution images. Besides, the non-linear operations in ViTs, e.g., LayerNorm (LN), GELU \cite{Hendrycks2016GaussianEL}, and especially Softmax, are hardware unfriendly and quantization sensitive \cite{Lin2021FQViTPQ, Yuan2021PTQ4ViTPQ}, hindering ViTs' achievable task accuracy and hardware efficiency.

\begin{figure}[htbp]
\centerline{\includegraphics[width=0.48\textwidth]{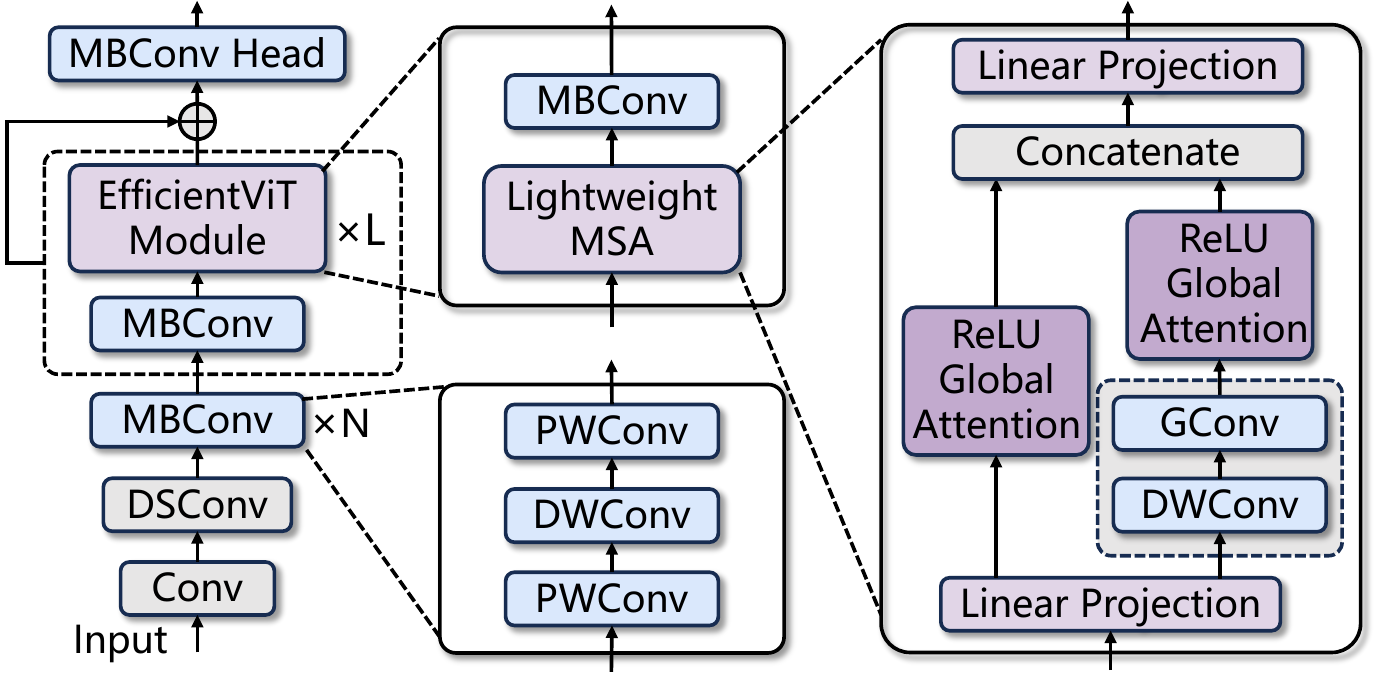}}
\caption{The macro architecture of EfficientViT \cite{cai2022efficientvit}. Each MBConv consists of two pointwise convolutions (PWConvs) separated by a depthwise convolution (DWConv). Besides, the key component of the EfficientViT module is the lightweight Multi-Scale Attention (MSA).}
\label{fig:efficientvit}
\vspace{-1.5em}
\end{figure}

To promote ViTs' deployment, various efforts have been devoted to the development of efficient ViTs \cite{Wang2021PyramidVT, pvtv2, cai2022efficientvit, Han2023FLattenTV}, which replace the vanilla computational-intensive self-attention in standard ViTs with more efficient alternatives that exhibit linear computational complexity.
However, it has been widely demonstrated that
the simplification of the attention mechanism inevitably results in a reduction in local feature extraction capabilities.
This limitation necessitates the incorporation of supplementary components such as convolutions \cite{cai2022efficientvit, Han2023FLattenTV}, yielding hybrid architectures for efficient ViTs that integrate both convolutions and Transformer blocks.
Particularly, the state-of-the-art (SOTA) efficient ViT, dubbed EfficientViT \cite{cai2022efficientvit}, can achieve higher accuracy than Swin-T \cite{Liu2021SwinTH} (by $+1.4\%$) and DeiT \cite{deit} (by $+2.9\%$) with a comparable number of parameters. 
As illustrated in Fig. \ref{fig:efficientvit}, EfficientViT features a Convolution-Transformer hybrid architecture, primarily comprising MBConvs \cite{Sandler2018MobileNetV2IR} and EfficientViT Modules. The latter includes a Softmax-free and lightweight Multi-Scale Attention (MSA), aiming to enhance both hardware efficiency and representation capability. 
Besides, EfficientViT also replaces vanilla LN and GELU in standard ViTs with hardware-friendly BatchNorm (BN) and Hardswish \cite{Howard2019SearchingFM}, respectively.


\shh{
Despite EfficientViT's effectiveness, existing accelerators \cite{Sun2022VAQFFA, You2022ViTCoDVT, Wang2022ViAAN, Li2022AutoViTAccAF} are mainly dedicated to standard ViTs \cite{vit, deit} and not directly applicable to accelerate EfficientViT.
To fully unleash its hardware benefit potential, it is highly desired to develop a dedicated accelerator for EffieicientViT, which, however, poses challenges due to its dynamic workloads and high-intensity memory access demands.
Particularly, EfficientViT involves various operation types, including lightweight convolutions (i.e., MBConvs) with different kernel sizes, strides, and feature dimensions, as well as the lightweight attention (i.e., MSA), which exhibits distinct computational patterns compared to the vanilla self-attention in standard ViTs. 
Moreover, the aforementioned lightweight components in EfficientViT exhibit reduced computing parallelism and fewer data reuse opportunities than their standard counterparts, yielding either high memory bandwidth requirement or low computation resource utilization.
Thus, in this paper, we present an FPGA-based accelerator for EfficientViT to tackle these challenges. The main contributions are summarized as follows.}
\begin{itemize}
\item \shh{A reconfigurable architecture is designed to efficiently support various operation types in the Convolution-Transformer hybrid architecture of EfficientViT, including lightweight convolutions and lightweight attention.} 
\item \shh{A novel time-multiplexed and pipelined dataflow is proposed to fuse computations among adjacent lightweight convolutions and computations within lightweight attention, dramatically boosting computing resource utilization while easing bandwidth requirements.}
\item \wdm{Based on optimizations of both computation and communication, an accelerator dedicated to EfficientViT is developed. It is implemented on the Xilinx ZCU102 FPGA platform at $200$MHz and achieves up to $780.2$ GOPS in throughput and $105.1$ GOPS/W in energy efficiency.}
\end{itemize}

\section{Structure of EfficientViT}\label{B}

\begin{figure}[tbp]
\centerline{\includegraphics[width=0.48\textwidth]{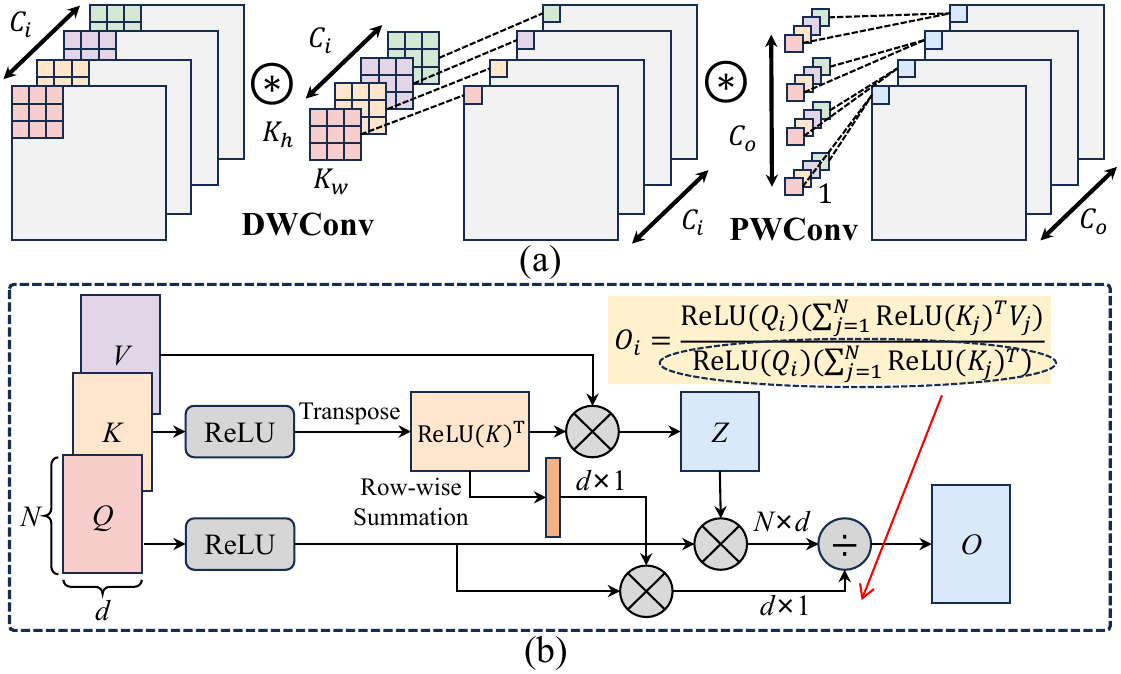}}  \vspace{-0.7em}
\caption{(a) DSConv: Depthwise Convolution followed by Pointwise Convolution. (b) The computation flow of ReLU-based global attention in EfficientViT.}
\label{fig:relu-attention} \vspace{-1.2em}
\end{figure}

As depicted in Fig. \ref{fig:efficientvit},
EfficientViT \cite{cai2022efficientvit} has an input stem of a generic convolution (Conv) followed by a DSConv layer, which is a combination of a depth-wise convolution (DWConv in Fig. \ref{fig:relu-attention}(a) left) and a point-wise convolution (PWConv in Fig. \ref{fig:relu-attention}(a) right). 
After that, two key types of blocks are involved in EfficientViT: the \textbf{MBConv} \cite{Sandler2018MobileNetV2IR} and the \textbf{EfficientViT Module}.   
The \underline{MBConv} features two PWConvs separated by a DWConv. Each layer is followed by BatchNorm (BN) and Hardswish activation \cite{Howard2019SearchingFM} (except the final PWConv).
Notably, BN can be implemented via $1\times1$ convolutions, which can be integrated into preceding convolutions to facilitate quantization and acceleration \cite{Zhang2022WSQAdderNetEW}.
In addition, each \underline{EfficientViT Module} comprises a {lightweight Multi-Scale Attention (MSA)} and an {MBConv} to separately extract context and local information.
In the MSA, inputs are projected to produce query/key/value ($Q/K/V$). Then, they are processed by lightweight convolutions to obtain multi-scale tokens, which further undergo ReLU-based global attention. Finally, the results are concatenated and projected to generate the final outputs.
As illustrated in Fig. \ref{fig:relu-attention}(b), the ReLU-based global attention transforms the similarity function in the original Softmax-based attention (i.e., $\text{Exp}(QK^T/\sqrt{d})$) into $\text{ReLU}(Q)\text{ReLU}(K)^T$, thus not only eliminating the need for Softmax but also achieving linear computational complexity by utilizing the associative property of matrix multiplication.





\section{Proposed Hardware Design}\label{C}



As discussed above, there are four main types of operations in the backbone of EfficientViT: generic Convs, PWConvs, DWConvs, and matrix multiplications (MatMuls).
As MatMuls can be treated as PWConvs with large batch sizes, an efficient hardware architecture that can effectively handle MSA and various types of convolutions is highly desired.
Additionally, lightweight operations (i.e., PWConvs/DWConvs/MSA) in EfficientViT features reduced computing parallelism and fewer data reuse opportunities, calling for an effective dataflow to enhance hardware utilization and ease bandwidth requirements.

\subsection{Multipliers and Adder-Trees Design Paradigm}\label{C1}
Convolutions in neural networks can be fundamentally decomposed into a series of multiplication and addition computations. 
Hence, parallelized hardware architectures incorporating multipliers and adder-trees (MAT) offer a straightforward and efficient solution for 
\shh{generic Convs and PWConvs (which are essentially generic Convs with $1$$\times$$1$ kernels)}.
Specifically,
\shh{inputs or weights can be broadcast to all MATs to facilitate data reuse.}
\shh{Within each MAT, multiple multipliers are responsible for generating partial sums along the input channel dimension in parallel, which are then added (via the adder tree) and accumulated to obtain the final output.}


\begin{figure*}[tbp]
\centerline{\includegraphics[width=1.0\textwidth]{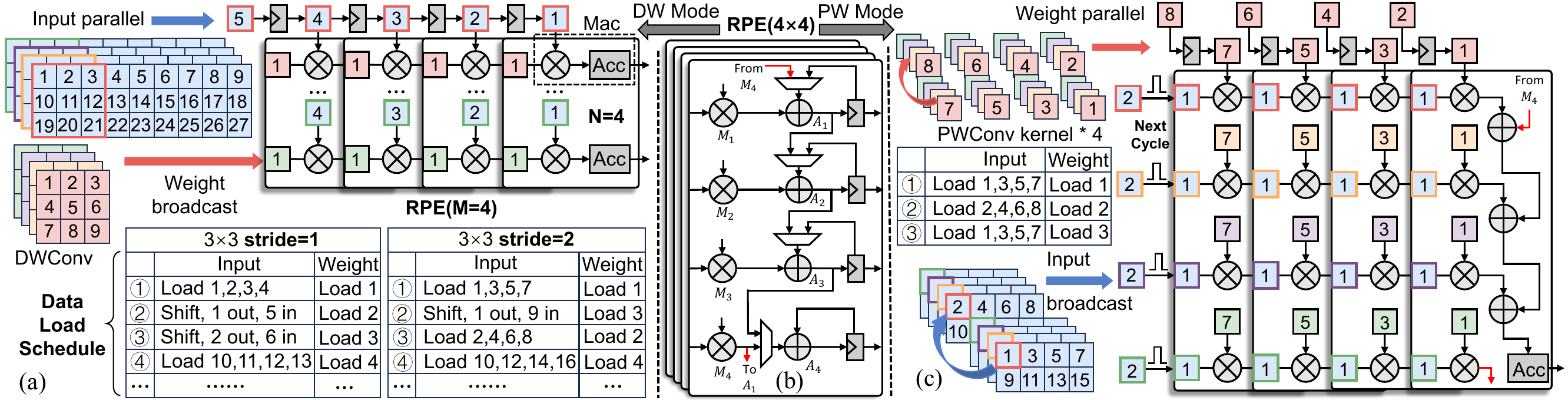}} \vspace{-0.7em}
\caption{(a) RPE works in DW Mode. (b) The micro-architecture of proposed Reconfigurable Processing Element (RPE). (c) RPE works in PW Mode.}
\label{fig:reconfig-pe} \vspace{-1.2em}
\end{figure*}

\shh{Despite its effectiveness, MAT-based architecture cannot effectively support DWConvs in EfficientViT due to its fixed structure and dataflow.
Particularly, DWConvs handle each input channel separately, thus only partial sums within the the same sliding window can be summed and accumulated. This constrains the achievable parallelism within each MAT to DW's kernel size, limiting MAT's flexibility for supporting various kernel sizes as well as its scalability to a large scale.}
\shh{Moreover, DWConvs with different kernel sizes and strides yield distinct overlap patterns between adjacent sliding windows when conducting convolutions, resulting in significant buffer overheads or complex memory management to support the generation of consecutive output pixels within MATs\cite{Yu2020LightOPUAF}.}

\subsection{Reconfigurable Architecture Design} \label{C2}
\shh{To boost flexibility, we develop a reconfigurable processing element (RPE) architecture to efficiently support various types of convolutions in EfficientViT. As depicted in Fig. \ref{fig:reconfig-pe} (b), RPE contains $M$ PE lines, each has $N$ multiplication-accumulation (MAC) units in parallel. It can be reconfigured to operate in both DW mode and PW mode to support DWConvs and PWConvs/generic Convs, respectively:}

\subsubsection{DW Mode}
\shh{When RPE works in the DW Mode, partial sums within each sliding window are accumulated within each MAC, which is an \textbf{self-accumulation} dataflow.}
\shh{Fig. \ref{fig:reconfig-pe}(a) shows an example of executing of a $k\times k$ ($k=3$ here) DWConv with stride $=1$ on the RPE with $M=4$.}
\shh{In the first cycle, inputs $a_1$ to $a_M$ are read and individually transferred to top MACs of different PE lines,
with weight $w_1$ loaded and broadcast.
In the second clock cycle, input data are right-shifted along registers with $a_1$ dequeued and $a_{M+1}$ enqueued. Weight is updated to $w_2$. 
Next cycle processes $a_3$$\sim$$a_{M+2}$ and $w_3$.
After $k$ cycles, the computation moves to the next row of the input feature map, following the same pattern. 
When $k$ rows of data are processed, the output $o_1$ to $o_M$ for DWConv can be obtained from the top MACs of $M$ PE lines via self-accumulation. 
The $N$ MACs within each PE line can conduct computations for $N$ output channels in parallel.}


\shh{
For $k\times k$ DWConv with a stride of 2, 
overlaps among consecutive sliding windows are spaced instead of successive. Thus, odd-column-indexed pixels within each row of the input feature map are first read and right-shifted by cycles, followed by the even-column-indexed pixels. Weights are also broadcast following the same ``first odd, then even" order to accommodate this modified computation scheme.
}

\subsubsection{PW Mode}
\shh{When RPE works in the PW mode to support both PWConvs and generic Convs, it performs a similar functionality as the MAT-based architecture. Specifically, as shown in Fig. \ref{fig:reconfig-pe} (c), partial sums along the input channel are computed via $N$ multipliers within each PE line and then accumulated down-forward the PE line, which is \textbf{down-forward accumulation} dataflow. This implies that the parallelism within each PE line is along the input channel dimension to leverage the partial sum reuse opportunity.
Besides, inputs can be broadcast among $M$ PE lines to exploit input reuse.
}


\subsection{Overall Architecture of Proposed Accelerator} \label{c3}
\begin{figure}[tbp]
\centerline{\includegraphics[width=0.48\textwidth]{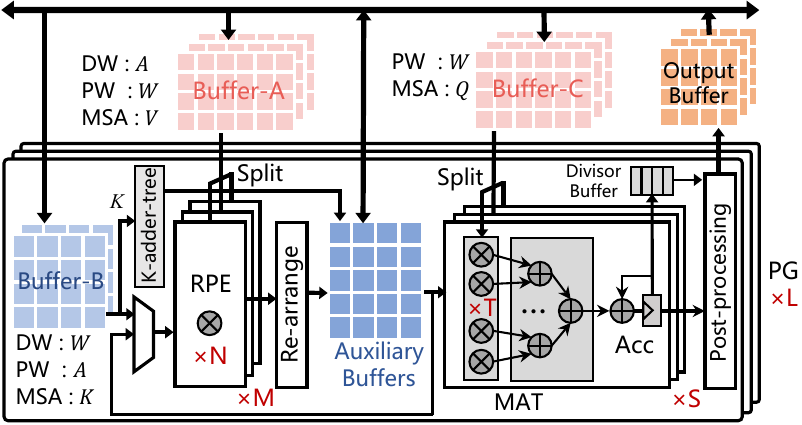}} \vspace{-0.7em}
\caption{The overall architecture of our accelerator. The RPE engine is composed of $M$ PE lines with $M\times N$ multipliers, and the MAT engine is configured as $S$ MATs with $S\times T$ multipliers.} 
\label{fig:accelerator} 
\vspace{-0.8em}
\end{figure}

Considering MAT's efficiency in executing the dominant PWConvs in EfficientViT and RPE's flexibility in supporting various operation types, we propose our architecture in Fig. \ref{fig:accelerator} incorporating both components to marry the best of both designs.
Particularly, our accelerator mainly comprises multiple on-chip buffers and $L$ parallel processing groups (PGs), each containing an RPE engine and a MAT engine.
The RPE engine can flexibly process DWConvs, PWConvs, generic Convs, and MatMuls, while the MAT engine is responsible for efficiently executing the latter three.

\begin{figure}[tbp]
\centerline{\includegraphics[width=0.43\textwidth]{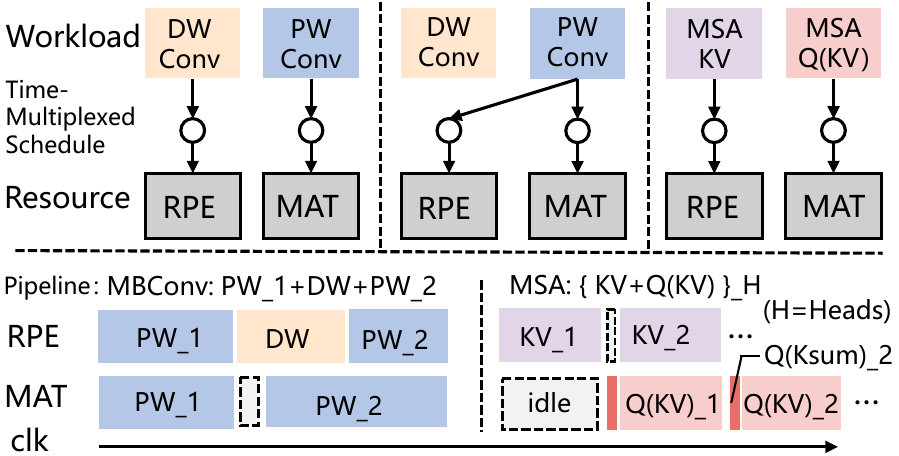}} \vspace{-0.8em}
\caption{The proposed time-multiplexed and pipelined (TMP) dataflow.}
\label{fig:schedule} \vspace{-1.6em}
\end{figure}

As illustrated in Fig. \ref{fig:accelerator}, buffers $A$, $B$, and $C$ are used to cache various types of data, including weight $W$, input $A$, as well as query $Q$, key $K$, and value $V$ within MSA.
During computation, data from buffers A and C are transmitted to all PGs. Then, they are split and separately sent to $M$ PE lines and $S$ MAT lines within each PG.
Besides, data read from the internal buffer B can be broadcast to $M$ PE lines.
The auxiliary buffers can connect the RPE and MAT engines and can also directly communicate with the off-chip DRAM.

\shh{Additionally, as shown in Fig. \ref{fig:relu-attention} (b),
in addition to MatMuls, MSA also involves row-wise summations and divisions, thus our accelerator also integrates auxiliary K-adder-tree and dividers to accommodate MSA's computation.}




\subsection{Time-Multiplexed and Pipelined Dataflow}


\shh{To enhance PE utilization and reduce data costs, we further equip our accelerator with a time-multiplexed and pipelined (TMP) dataflow to facilitate both the 
\underline{(1)} \textbf{inter-layer} fusion in MBConvs and \underline{(2)} \textbf{intra-layer} fusion for computations within MSA.
Firstly, considering that DWConvs can only be executed on the RPE engine and they are always followed by PWConvs in MBConvs of EfficientViT, we thus fuse DWConvs with their subsequent PWConvs.
As depicted in Fig. \ref{fig:schedule}, when DWConv is executed on the RPE engine, partial sums can be cached in the auxiliary buffer, then generated outputs can be immediately passed to the idle MAT engine to serve as input for its subsequent PWConv. 
As DWConvs involve fewer computations than PWConvs, when the RPE engine finishes processing DWConv, it can join the computation of the concurrent PWConv to boost hardware utilization.}

\shh{Additionally, as for intra-layer fusion for MSA, MatMuls of $Z=\text{ReLU}(K^T)\cdot V$ and $\text{ReLU}(Q)\cdot Z$ can also be pipeline-executed on two engines.
Specifically, when $\text{ReLU}(K^T)$ is loaded from Buffer $B$ to the RPE engine for conducting MatMuls with $V$, K-adder-tree can simultaneously perform row-wise summations of $\text{ReLU}(K^T)$ to obtain $\text{ReLU}(K)^T_\text{sum}$.
The resultant outputs, i.e., $\text{ReLU}(K)^T_\text{sum}$ and $Z$ are saved in the auxiliary buffers and then broadcast to the MAT engine to sequentially conduct multiplications with $Q$, generating divisors and dividends of MSA, respectively.
During this process, the pre-generated divisors are temporarily saved in a small divisor buffer.
Once dividends are computed by the MAT engine, they can be divided by the previously saved divisors via dividers in the post-processing module (Fig. \ref{fig:accelerator}) to accomplish the final divisions of MSA.
}
\section{Experimental Results}\label{D}
\begin{table}[t]
\caption{FPGA Resource Utilization} \vspace{-1em}
\begin{center}
\begin{tabular}{c|c c c c }
\toprule[1pt] 
\rule{0pt}{8pt}
     & LUT  & FF & BRAM & DSP  \\
\hline 
\rule{0pt}{8pt}
Used & 104463  & 249473 & 160 & 1024 \\
\rule{0pt}{8pt}
Available & 274080 & 548160 & 912 & 2520\\
\rule{0pt}{8pt}
Utilization & 38.11\% & 45.51\% & 17.54\% & 40.63\% \\
\bottomrule[1pt] 
\end{tabular}
\label{tab1}
\end{center} \vspace{-1.0em}
\end{table}

\begin{figure}[t]
\centerline{\includegraphics[width=0.43\textwidth]{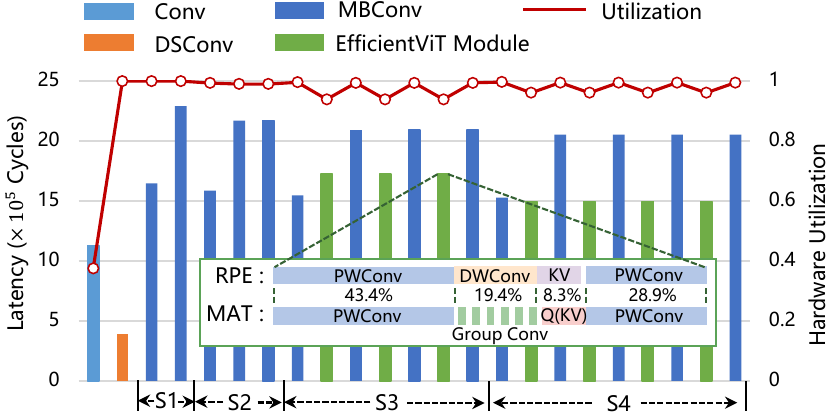}} \vspace{-0.6em}
\caption{The latency and hardware utilization evaluated on EfficientViT-B1, containing a generic Conv, a DSConv layer, and four stages (S1-S4).} 
\label{fig:latency} \vspace{-1em}
\end{figure}

\subsection{Experimental Setup}\label{D1}
Our accelerator is coded with Verilog, synthesized and implemented by Vivado Design Suite, and evaluated on Xilinx ZCU102 FPGA at 200-MHz frequency. The hardware resource of $(M\times N+S\times T)\times L$ is configured as $(8\times 8+8\times 8)\times 16$.
\shh{Each multiplier in both RPE and MAT engines can execute the $8\times 8$-bit fixed-point (FIX8) multiplication. Thus, to enhance DSP utilization, we adopt the SOTA DSP packaging method \cite{Xilinx-conv} to accommodate two $8\times 8$-bit multiplications within each DSP following Auto-ViT-Acc \cite{Li2022AutoViTAccAF} for fair comparisons.}
The resource consumption is reported in Table \ref{tab1}.

\subsection{Performance Analyses}
\shh{From Fig. \ref{fig:latency}, which is evaluated on the EfficientViT-B1 \cite{cai2022efficientvit} model, we can draw the following conclusions:}
\shh{\underline{(1)} As the input image has only $3$ channels and thus cannot be effectively mapped to our accelerator with high parallelism, this results in a low hardware utilization of $37.5\%$ when executing the first generic Conv on our design;
\underline{(2)} Similarly, the group Convs in MSA also have fewer input channel parallelism opportunities than PWConvs, yielding a slight utilization decrease here.
\underline{(3)} However, due to the effectiveness of our proposed TMP dataflow in fusing dominant computations among DWConvs and PWConvs as well as within MSA, the overall utilization is above $\mathbf{95}\%$, achieving a throughput of 780.2 GOPS and demonstrating our superiority.}

\vspace{-1em}

\begin{table}[ht]
  \caption{Comparisons with SOTA Works} 
  \vspace{-1em}
  \begin{center}
  \begin{threeparttable}
  \begin{tabular}{c| c| c| c| c}
  \toprule[1pt]
 & \makecell[c]{Efficient\\ViT\cite{cai2022efficientvit}}  &   \makecell[c]{ViA\cite{Wang2022ViAAN}}  & \makecell[c]{Auto-ViT-\\Acc\cite{Li2022AutoViTAccAF}}  &  \makecell[c]{Our work} \\
  \hline
  \rule{0pt}{8pt}
  \makecell[c]{Device}                  & \makecell[c]{CPU\tnote{*}} & \makecell[c]{Xilinx\\ Alveo U50} & \makecell[c]{Xilinx\\ ZCU102} & \makecell[c]{Xilinx\\ ZCU102}  \\
  \hline
  \rule{0pt}{8pt}
  \makecell[c]{Frequency (GHz)}               & \makecell[c]{1.8-3.0} & 0.3 & 0.15 & 0.2   \\
  \hline
  \rule{0pt}{8pt}
  \makecell[c]{Precision}                     & FP32 & FP16 & FIX8 & FIX8  \\
  \hline
  \rule{0pt}{8pt}
  \makecell[c]{DSP Used}                      & - & 2420  & 1936  & 1024  \\
  \hline
  \makecell[c]{Throughput\\ (GOPS)}           & 54.7 & 309.6  & 711.2  &  780.2  \\
  \hline
  \rule{0pt}{8pt}
  \makecell[c]{Power (W)}                     & 11 & 39 & 8.46 &  7.43   \\
  \hline
  \makecell[c]{Energy Efficiency\\ (GOPS/W)}  & 4.97 & 7.92 & 84.1 &  105.1  \\
  \hline
  \rule{0pt}{8pt}
  \makecell[c]{DSP Efficiency\\ (GOPS/DSP)}    & - & 0.13 & 0.37 &  0.76   \\
  \bottomrule[1pt] 
  \end{tabular}
  \begin{tablenotes}
    \footnotesize      
        \item[*] Qualcomm Snapdragon 8Gen1 CPU with 11W peak power consumption. 
  \end{tablenotes}
  \end{threeparttable}
  \label{tab2}
  \end{center}
  \vspace{-1.6em}
\end{table}

\vspace{-0.8em}
\subsection{Comparisons and Discussion}
\shh{To verify our accelerator when executing EfficientViT, we compare with prior works: EfficientViT-B1 \cite{cai2022efficientvit} measured on a mobile CPU with FP32 format, a SOTA Swin-Transformer \cite{Liu2021SwinTH} (also an efficient ViT) accelerator ViA\cite{Wang2022ViAAN} with FP16 format, and a standard ViT (DeiT) accelerator Auto-Vit-Acc \cite{Li2022AutoViTAccAF} with FIX8 precision.
From Table \ref{tab2}, we can see that:
\underline{(1)} Compared with EfficientViT on CPU, we can gain \textbf{14.3}$\times$ speedup and $\uparrow$\textbf{21.1}$\times$ energy efficiency;
\underline{(2)} Compared to ViA, our design achieves $\uparrow$2.0$\times$ throughput, $\uparrow$\textbf{13.3}$\times$ energy efficiency, and $\uparrow$\textbf{5.9}$\times$ DSP efficiency;
\underline{(3)} Although Auto-ViT-Acc consumes $1.9\times$ more DSP resources than us, we can offer $\uparrow$1.1$\times$ throughput, $\uparrow$$1.25\times$ energy efficiency, and $\uparrow$$2.1\times$ DSP efficiency, further validating our effectiveness.
}
\vspace{-0.6em}
\section{Conclusion}\label{E}
In this paper, we proposed an FPGA-based accelerator for Convolution-Transformer hybrid networks like EfficientViT. 
Specifically, we design a reconfigurable design to effectively support various types of convolutions and the Multi-Scale Attention (MSA).
Furthermore, we propose a time-multiplexed and pipelined dataflow to facilitate layer/computation fusions, boosting hardware utilization and minimizing bandwidth requirement.
Implemented results show that we can achieve up to 780.2 GOPS in throughput and 105.1 GOPS/W in energy efficiency, significantly outperforming prior works.



\bibliographystyle{ieeetr}
\bibliography{reference}

\vspace{12pt}

\end{document}